\documentclass[aps,prb,reprint,groupedaddress,showkeys]{revtex4-1}

\usepackage{graphicx}
\usepackage{dcolumn}
\usepackage{bm}

\begin{document}


\title{Polymorphism in elemental silicon: Probabilistic interpretation of the realizability of metastable structures}


\author{Eric Jones}
\email[]{ebjones@mymail.mines.edu}
\affiliation{Colorado School of Mines, Golden, Colorado 80401, USA}
\affiliation{National Renewable Energy Laboratory, Golden, Colorado 80401, USA}

\author{Vladan Stevanovi\'{c}}
\email[]{vstevano@mines.edu}
\affiliation{Colorado School of Mines, Golden, Colorado 80401, USA}
\affiliation{National Renewable Energy Laboratory, Golden, Colorado 80401, USA}


\date{\today}

\begin{abstract}
With few systems of technological interest having been studied as extensively as elemental silicon, there currently exists a wide disparity between the number of predicted low-energy silicon polymorphs and those, which have been experimentally realized as metastable at ambient conditions. We put forward an explanation for this disparity wherein the likelihood of formation of a given polymorph under near-equilibrium conditions can be estimated on the basis of mean field isothermal-isobaric $(N, p, T)$ ensemble statistics. The probability that a polymorph will be experimentally realized is shown to depend upon both the hypervolume of that structure's potential energy basin of attraction and a Boltzmann factor weight containing the polymorph's potential enthalpy per particle. Both attributes are calculated using density functional theory relaxations of randomly generated initial structures. We find that the metastable polymorphism displayed by silicon can be accounted for using this framework to the exclusion of a very large number of other low-energy structures.
\end{abstract}

\pacs{}
\keywords{Silicon, Polymorph, Allotrope, Crystal Structure Prediction, Statistical Mechanics, Experimental Realizability}

\maketitle


\section{\label{sec:exp_real} Introduction}

Driven largely by the success of its monocrystalline ground state as the cornerstone of modern semiconductor technology, elemental silicon continues to see major efforts both to understand and to consider applications for the prolific polymorphism it displays \cite{doi:10.1063/1.4962984}. While equilibrium methods for studying Si polymorphism, such as application of pressure via diamond anvil cell (DAC), have been employed for over 50 years \cite{Jamieson762}, recent advances in discovering new polymorphs have relied upon unconventional synthesis methods such as ultrafast laser-induced confined microexplosion\cite{rapp2015experimental} and high pressure precursor routes \cite{Kim2015}. Theoretical activity geared towards predicting new silicon polymorphs has kept pace, and as is summarized by Haberl et al.\cite{doi:10.1063/1.4962984}, predictions of hitherto un-sythesized metastable structures now numbers in the many dozens, with proximity in energy to the ground state as the main criterion used to judge experimental realizability.
%
\begin{figure}
\includegraphics[width=\linewidth]{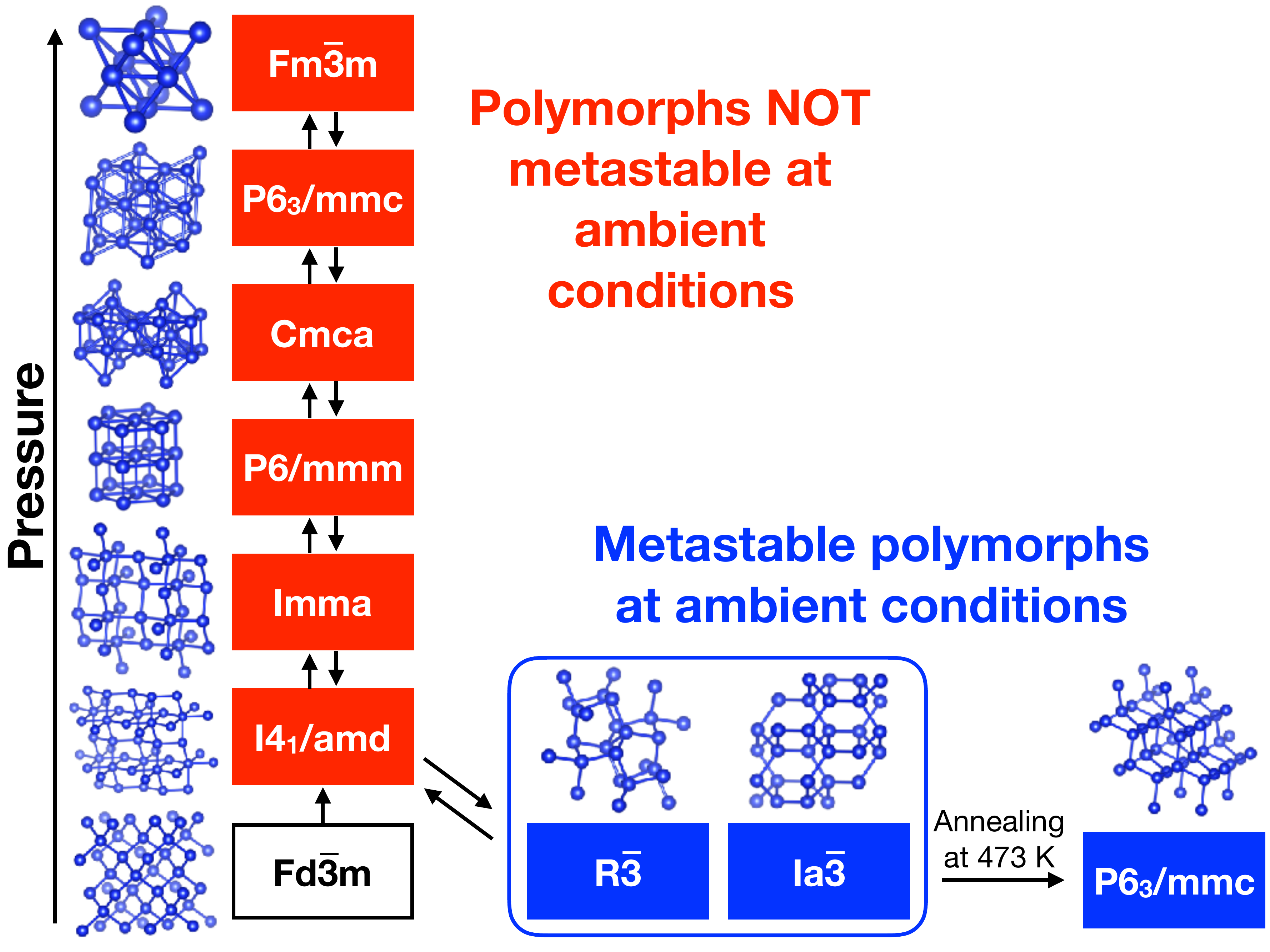}
\caption{\label{fig:eq_polys} Polymorphs of silicon synthesized under equilibrium conditions. Bidirectional arrows denote reversible, pressure-induced phase transformations while a unidirectional arrow denotes an irreversible transformation. Polymorphs labeled with red boxes are unstable at ambient conditions, while those labeled with blue are metastable at ambient conditions, and the white box represents the cubic diamond ground state.}
\end{figure}
%
The experimental polymorph spectrum of silicon generated by equilibrium and nearer-to-equilibrium methods has been thoroughly characterized over a wide range of temperatures and pressures and is schematized in Fig. \ref{fig:eq_polys}. Note that on first reference, equivalent naming conventions will be placed in round brackets next to the colloquial name. Thereafter, polymorphs will be referenced by their space group assignment. Upon compression in a DAC, the diamond ground state (dc-Si, Si-I, Fd$\bar{3}m$) metalizes into the beta tin phase ($\beta$-Sn-Si, Si-II, I4$_1$/amd) at 11.7 GPa \cite{Jamieson762} before passing through two closely related higher pressure phases: orthorhombic Imma (Si-XI, Imma) at 13.2 GPa \cite{PhysRevB.47.8337, PhysRevB.50.739} and simple hexagonal (sh-Si, Si-V, P6/mmm) at 15.3 GPa \cite{PhysRevB.47.8337, OLIJNYK1984137}. This series of phase transitions is reversible upon slow decompression back to the I4$_1$/amd phase, but upon full decompression at ambient temperatures, the body-centered structure with 8 atoms in its primitive unit cell (bc8-Si, Si-III, Ia$\bar{3}$) is recovered along with trace remnants of the rhombohedral 8 atom polymorph (r8-Si, Si-XII, R$\bar{3}$) \cite{MINOMURA1962451, Wentorf338}. Pressure release from an I4$_1$/amd phase obtained by indentation loading produces a higher fraction of R$\bar{3}$ to Ia$\bar{3}$ in the resulting metastable mixed phase. Hence, it appears that the formation of R$\bar{3}$ as a metastable state is concomitant to the formation of Ia$\bar{3}$ as well \cite{PhysRevB.83.075316}. By annealing Ia$\bar{3}$ produced by DAC from room temperature to 473 K, it transforms into a third metastable polymorph, the hexagonal diamond, lonsdaleite analogue (hd-Si, Si-IV, $P6_3/mmc$) \cite{0268-1242-4-4-029}. Meanwhile, \textit{ex situ} annealing of the mixed Ia$\bar{3}$/R$\bar{3}$ phase produced by indentation loading to 473 K results in an as yet poorly characterized phase (Si-XIII, P4$_1$2$_1$2), which has only been reported to occur as a phase mixture \cite{doi:10.1063/1.3124366}.

Our emphasis here will be on metastable polymorphs of silicon. For completeness however, we mention that upon further compression, P6/mmm transforms reversibly into orthorhombic Cmca (Si-VI, Cmca) at 38 GPa \cite{PhysRevLett.82.1197}, hexagonal close pack (hcp-Si, Si-VII, P6$_3$/mmc) at 42 GPa \cite{OLIJNYK1984137, PhysRevLett.82.1197} (any future reference to P6$_3$/mmc will refer to the metastable lonsdaleite phase), and face-centered cubic (fcc-Si, Si-X, Fm$\bar{3}$m) at 79 GPa \cite{PhysRevB.41.12021}.  Two further metastable phases result exclusively upon \textit{rapid} pressure release from 14.8 GPa (Si-VIII, P4$_1$2$_1$) and 12 GPa (Si-IX, P4$_2$2) \cite{ZHAO1986679}, but neither has seen a full characterization of its crystal structure. Therefore, the polymorphs of silicon, which are both synthesized under near equilibrium conditions and metastable at ambient conditions are plausibly considered to be those denoted by blue boxes in Fig. \ref{fig:eq_polys}: R$\bar{3}$, Ia$\bar{3}$, and P6$_3$/mmc.

That this number of metastable, equilibrium-formed polymorphs is so few compared to the multitude, which are energetically competitive with the ground state in the silicon potential energy surface (PES) \cite{PhysRevB.86.121204}, motivates the development of some organizing principle by which candidate metastable polymorphs can be assessed for their likelihood of experimental realization. As recognized by authors such as Stillinger \cite{stillinger2015energy}, a natural choice of formalism is statistical mechanics, since it simultaneously allows for a landscape of microstates to be meaningfully related to the macroscopic properties of a system and explicitly takes thermodynamic inputs such as temperature and pressure as parameters of the theory. In addition, it has recently been shown that the experimental polymorphism displayed by MgO, ZnO, and SnO$_2$ can be largely accounted for by considering the hypervolume of a polymorph's basin of attraction \cite{PhysRevLett.116.075503}. Since these three systems also possess very large numbers of potential polymorphs uncovered by computation, the introduction of basin hypervolume as an extra constraint in the prediction process was shown to be necessary in order to accurately select those polymorphs, which appear experimentally. This basin hypervolume constraint was given a probabilistic interpretation, which further suggests a fully probabilistic framework within which to discuss the experimental realizability of polymorphs.

Of course, any complete analysis of metastable polymorphism for a given compound must also involve determination of kinetic barriers between candidate polymorphs. Our model serves as an initial screening process so that the number of candidate polymorphs between which kinetic barriers must be calculated can be drastically reduced in order to make more efficient the selection of realizable, kinetically stable polymorphs from a multitude of possible low-energy structures. True metastable polymorphs from the model-selected set will then be those that are also kinetically stable. It is to the construction of this model that we now turn.
\section{\label{sec:stat_mech}Polymorph Realizability from Isothermal-Isobaric Ensemble Statistics}
%
\begin{figure}
\includegraphics[width=\linewidth]{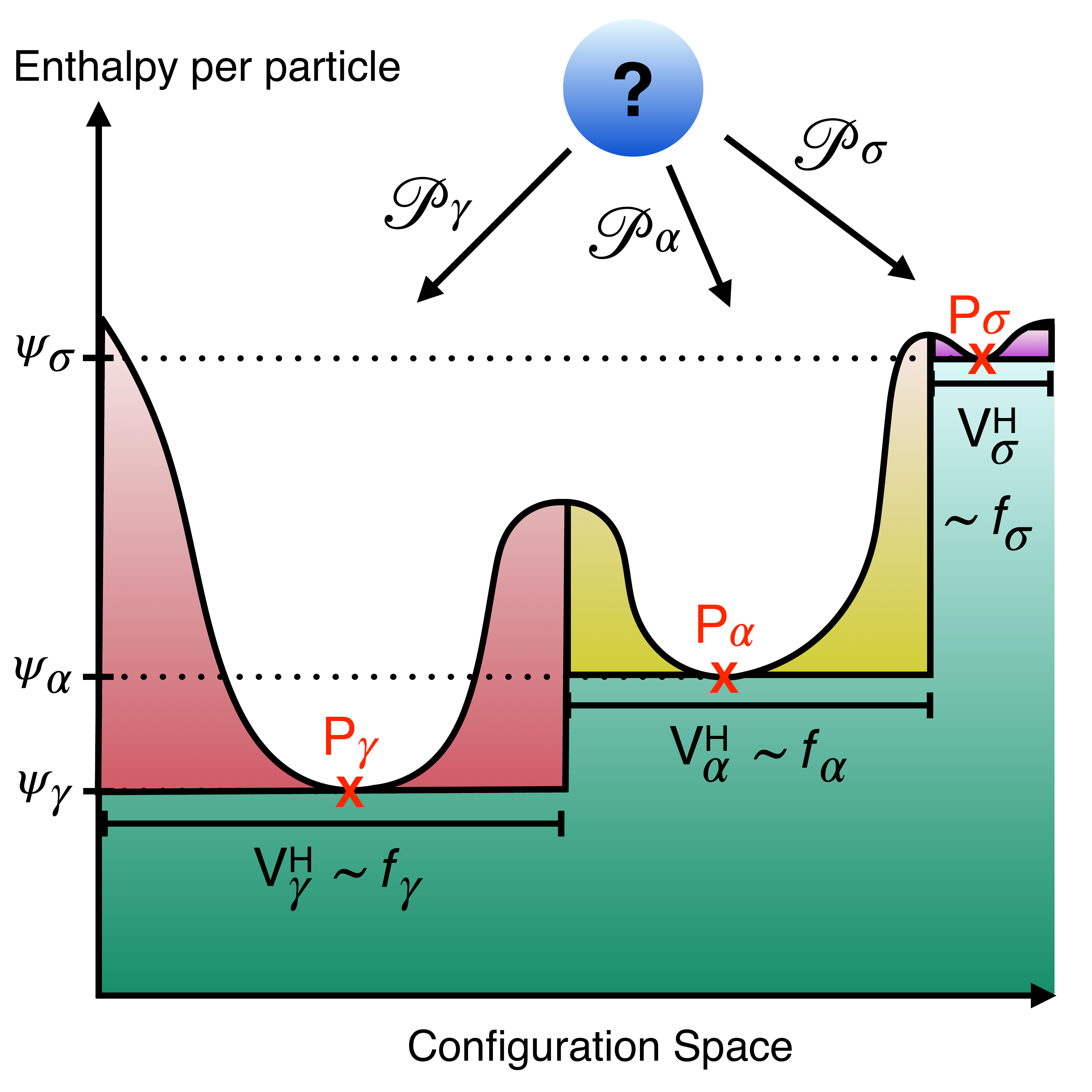}
\caption{\label{fig:ion_choosing} Cartoon model of how an ion in a bulk material decides which structural configuration to become a part of. The ion is more likely to find minima of basins with large hypervolume ($V^H_{\gamma}$), which is proportional to a structure's frequency of occurrence ($f_{\gamma}$). Of the minima it finds, it will try to join the corresponding structure, which minimizes its enthalpy ($\psi_{\gamma}$). The interplay between these two effects is neatly accounted for by the probability of occurrence $\mathcal{P}_{\gamma} = f_{\gamma} e^{- \beta \psi_{\gamma}}$.}
\end{figure}
In order to develop a model, which accurately accounts for silicon's experimental behavior, we consider the perspective of an ion deciding which of a number of nucleated structures it would most like to join, represented by the matte blue circle in Fig. \ref{fig:ion_choosing}. Note that this mean field, single-particle picture is already tacitly assumed in much of the polymorph prediction community since activation barriers and total energies are rarely cited as a function of all N particles simulated \cite{Khaliullin2011} but rather as energy per atom. We therefore begin our analysis with the partition function for a single particle immersed in an N-particle ionic configuration (corresponding to the N particles in a simulated unit cell) with 3N coordinates $\bm{r}$, periodic boundary conditions, average volume per atom $v$, mean field potential energy per particle $\phi(\bm{r}, v)$, and held under constant temperature $(T)$ and pressure $(p)$.
%
\begin{equation}\label{eq:real_one}
\Xi_1(N,p,T) = \frac{1}{V_0 \Lambda_T^3} \int dv \int d\bm{r} e^{-\beta (\phi(\bm{r},v)+pv)}
\end{equation}
%
$\Lambda_T = h/\sqrt{2 \pi M k_B T}$ is the mean thermal wavelength produced by the separable momentum integral, $V_0$ is a constant with units of volume to make the right-hand side dimensionless, and $\beta=1/k_BT$. The mean-field treatment of the interactions between the atoms in a solid we find appropriate because it allows elimination of the dependence on the number of atoms $N$ from the partition function. Also, it is important to note that the Born--von Karman boundary conditions are assumed, that is, the finite size effects such as the surface and interface energies that are relevant for the nucleation and growth of different phases are not accounted for in this discussion. Meanwhile, the integrand in Eq. \ref{eq:real_one} can be interpreted as the probability density for a single ion to find itself as part of an ionic configuration $\bm{R}=(\bm{r},v)$ with mean field potential enthalpy per particle $\psi(\bm{R})=\phi(\bm{r},v)+pv$.

Let $\alpha$ index structures $P_{\alpha}$, which are local potential enthalpy minima, indicated in bright red in Fig. \ref{fig:ion_choosing}. We also introduce $B_{\alpha}$ as the basin of attraction leading to $P_{\alpha}$ with attendant intra-basin coordinates $\bm{R}_{\alpha}$. The configurational integral can then be carved up into a sum of integrals over basins
%
\begin{equation}\label{eq:my_seven}
\Xi_1(N, p, T) = \frac{1}{V_0 \Lambda^{3}_T} \sum_{\alpha} \int_{B_{\alpha}} d\bm{R}_{\alpha} e^{-\beta \psi \big(\bm{R}_{\alpha} \big)}\,.
\end{equation}
%
To lowest order, we approximate each basin's contribution to the integrand by $\psi(\bm{R}_{\alpha}) \rightarrow \psi_{\alpha}$, the value of the potential enthalpy per particle evaluated at the basin minimum. As is represented in Fig. \ref{fig:ion_choosing} by the matte red, yellow, and purple coloration, this approximation amounts to a flattening of each basin into a ``square'' well. The integral then yields the hypervolume $(V^H_{\alpha})$ of the square well that maps to $P_{\alpha}$, the well's local minimum. With these modifications, an estimate for the relative probability for an ion to choose between two nascent polymorphs $P_{\alpha}$ and $P_{\gamma}$ results.
%
\begin{equation}\label{eq:my_eight}
\frac{\mathcal{P}_{\alpha}}{\mathcal{P}_{\gamma}} = \Bigg( \frac{V^H_{\alpha}}{V^H_{\gamma}} \Bigg) e^{-\beta (\psi_{\alpha} - \psi_{\gamma})}
\end{equation} 
%
The Boltzmann factor in Eq. \ref{eq:my_eight} exponentially suppresses states, which are far away from the thermodynamically most favorable state at a given $(T, p)$, while the ratio $V^H_{\alpha}/V^H_{\gamma}$ selects which states around the global enthalpy minimum are more likely to form.

In practice, interest in metastable states, which are energetically competitive with the ground state has typically dictated that consideration of the $(T=0,p=0)$ potential energy surface is sufficient in order to identify candidate polymorphs for synthesis \cite{doi:10.1063/1.4962984}. Under these circumstances, $\psi_{\alpha} \rightarrow \phi_{\alpha} =  \Phi_{\alpha}/N$, the potential energy per particle. However, simply because a state is metastable at ambient conditions does not mean that it necessarily forms initially under those conditions. We therefore would like an estimate for the potential enthalpy proper ($\psi_{\alpha}$) such as to allow for varying synthesis conditions, but one that is also not too dissimilar to the PES at $(T=0,p=0)$. One can recover such an estimate for the enthalpy per particle by using the energy and volume per ion evaluated at $(T=0, p=0)$, $\psi^{approx.}_{\alpha} = \phi_{\alpha} + pv^{approx.}_{\alpha}$ (again, $p$ is the pressure to be inserted as a parameter). A relevant assumption for our analysis then becomes that the ratios $V^H_{\alpha}/V^H_{\gamma}$ remain mostly unaltered for states not too far away from the ground state. This assumption is justified on the basis that the potential enthalpy surface at low to moderate pressures is expected to be an adiabatic deformation of the potential energy surface at zero pressure \cite{verma1966polymorphism}. $\psi^{approx.}_{\alpha}$ allows us to evaluate the Boltmann factor in Eq. \ref{eq:my_eight}.

Meanwhile, one can estimate the prefactor $V^H_{\alpha}/V^H_{\gamma}$ by populating the PES with $N^{RS}$ randomly generated structures. If $N_{\alpha}$ is the number of those structures, which are assignable to the local minimum $P_{\alpha}$ by virtue of being in the basin of attraction $B_{\alpha}$ then
%
\begin{equation}
\frac{V^H_{\alpha}}{V^H_{\gamma}} = \lim_{N^{RS} \rightarrow \infty} \Bigg( \frac{N_{\alpha}}{N^{RS}} \Bigg) \bigg/ \Bigg( \frac{N_{\gamma}}{N^{RS}} \Bigg) \equiv \frac{f_{\alpha}}{f_{\gamma}},
\end{equation}
%
where the $f_{\alpha}$ are termed ``frequencies of occurrence'' \cite{PhysRevLett.116.075503}. The $\mathcal{P}_{\alpha} = f_{\alpha} \exp({-\beta \psi^{approx.}_{\alpha}})$ are termed ``probabilities of occurrence'' and are the main objects of interest in this paper.

Finally, the connection between single particle statistics and the thermodynamics of a bulk solid can be made through the Gibbs free energy per particle.
%
\begin{equation}\label{eq:init_gibbs}
g = - \frac{1}{\beta} \ln \big( \sum_{\alpha} f_{\alpha} e^{- \beta \psi_{\alpha}} \big)
\end{equation}
%
If a given state  $\gamma \in \{\alpha\}$ is heavily favored in the above summation then to lowest order Eq. \ref{eq:init_gibbs} can be written
%
\begin{equation}\label{eq:g_two}
g \approx \psi_{\gamma} - \frac{1}{\beta} \ln f_{\gamma}.
\end{equation}
Hence under a given $(T,p)$, a resonant probability of occurrence of a given state drives the system to both minimize its enthalpy per particle and maximize its entropy per particle to $s = k_B \ln f_{\gamma}$, by choosing that state.

Since there are a number of approximations, which have been made in this section in order to estimate the contributions to the partition function in Eq. \ref{eq:real_one}, we will look for qualitative features in the probability distribution $\mathcal{P}_{\alpha}$ that might, under certain experimental conditions, lead to resonant probabilities of the type in Eq. \ref{eq:g_two}. To this end, we detail in Sec. \ref{sec:methods} how random structures are initialized on the PES, mapped to local minima, and sorted into structural equivalence classes. 
%
\section{\label{sec:methods}Computational Evaluation of Potential Enthalpy Surface Statistics}
%
\subsection{Random Structure Sampling}
%
In order to initialize random structures on the silicon PES, we utilize the previously developed procedure for doing so \cite{PhysRevLett.116.075503}. Each unit cell is specified by six parameters chosen randomly, the lengths of the unit cell's lattice vectors $a,b, c$ bounded between $0.6$ and $1.4$ in scaled units, and the angles between them $\alpha, \beta, \gamma$ bounded between $60^{\circ}$ and $140^{\circ}$. In fractional coordinates, ions are randomly distributed within the unit cell by first constructing a reciprocal lattice vector $\textbf{G}=n_1\textbf{g}_1+n_2\textbf{g}_2+n_3\textbf{g}_3$ with $n_1, n_2, n_3$ all between $4$ and $7$. This defines a plane wave $\cos(\textbf{G} \cdot \textbf{r})$.  At each plane wave minimum, silicon ions are randomly distributed into planes defined by $\textbf{G}$. In order to minimize instances where ions are randomly placed too close together, Gaussian probability distributions are centered on atoms, which have already been placed and subsequent ions are preferentially located in low probability regions on each superlattice plane. Once all of this has been done, a coordinate transformation is performed back into cartesian coordinates and the random structure is scaled such that the minimum distance between any two ions is no shorter than 1.8 {\AA}, which helps with convergence of subsequent DFT relaxations and sets the scale of the dynamics to the physical scale at which they occur. In order to ensure that the resulting statistics were robust under changes in this minimal bond length, sets of relaxations were performed where the minimal bond length was adjusted to as small as 1.4 {\AA} and as large as 2.2 {\AA}. No subsequent dependence of the sampling statistics was found.
%
\subsection{DFT Relaxations}
%
The mapping $\psi(\bm{R}_{\alpha}) \rightarrow \psi_{\alpha}$ depicted in Fig. \ref{fig:ion_choosing} is implemented by relaxing initialized random structures to nearby local minima via Density Functional Theory. We note that using this method, DFT relaxations can push a random structure initialized in one basin over a small energy barrier into an adjacent basin. Thus, it will be more accurate in the following to say that the $\alpha$ index funnels of attraction \cite{PhysRevLett.116.075503}. At each relaxation step the electronic ground state is computed for the ionic configuration at that step using the Perdew, Burke, Ernzerhof (PBE) form for the exchange correlation functional \cite{PhysRevLett.77.3865}. Valence electron behavior near ionic cores is accounted for using the projector augmented wave (PAW) method \cite{PhysRevB.50.17953}, and the simulation package used for calculations is the Vienna Ab initio simulation package (VASP) \cite{CMS.6.15}. At each step, ionic positions, unit cell volume and unit cell shape are all relaxed using the conjugate gradient algorithm \cite{Press:2007:NRE:1403886}. This process is continued until the total energy is converged to within $3$ meV/atom between successive iterations. In order to ensure correct numerical convergence, both volume and ionic relaxations are restarted four times and complemented by a self consistent DFT run afterwards. In addition, relaxations resulting in structures with residual forces above $10^{-4}$ eV/{\AA} and/or pressures above $3$ kbar are restarted in order to alleviate these residual forces and pressures. The pylada wrapper for high-throughput calculations was used in order to manage the computations and workflows \cite{pylada}.
%
\subsection{Structure Comparison}
%
Once initialized random structures are successfully relaxed to local minima, the resulting structures are compared and sorted into classes of structural equivalence according to four attributes: ({\it i}) common total energy to within $10$ meV/atom, ({\it ii}) identical space groups to within a tolerance of $0.3$ {\AA}, ({\it iii}) volume per atom match to within a tolerance of $1 \%$, and ({\it iv}) coordination of up to the fourth nearest neighbor to within a tolerance of $0.2$ {\AA}. The total energy per atom and volume per atom of each equivalence class is then used to compute the approximate enthalpy per particle of that equivalence class ($\psi^{approx.}_{\alpha}$), and the ratio of the equivalence class magnitude to the total number of relaxed structures results in the frequency of occurrence ($f_{\alpha} = N_{\alpha}/N^{RS}$).
%
\section{\label{sec:results}Results and Discussion}
%
\begin{figure*}
\includegraphics[width=\linewidth]{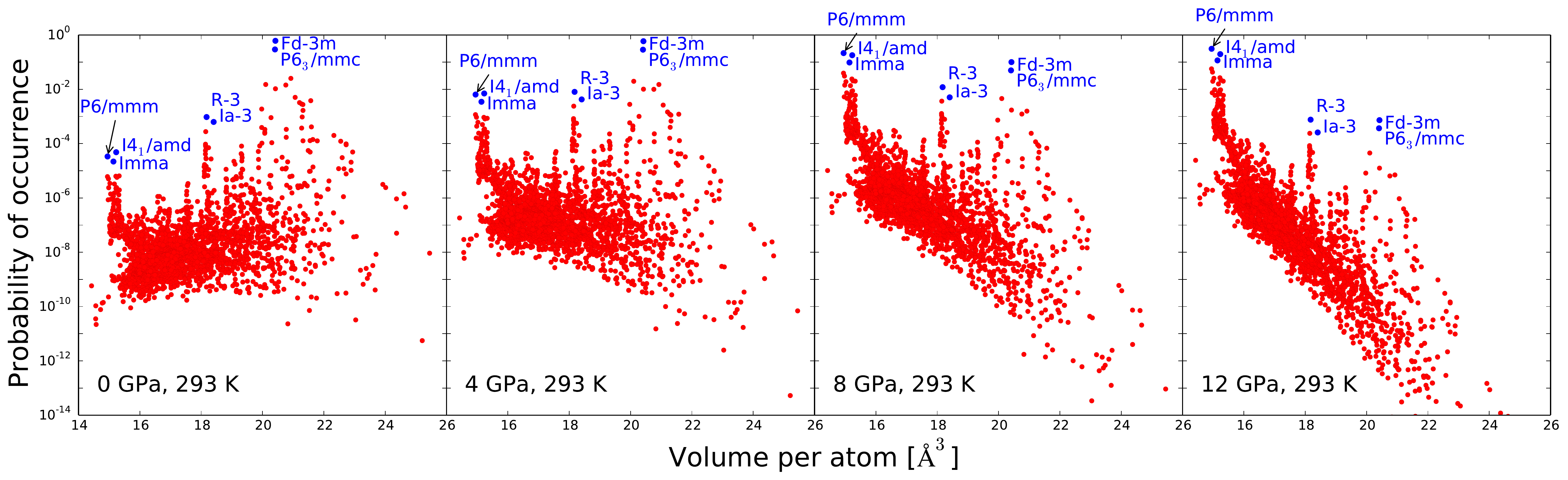}
\caption{\label{fig:press_deps} Plot of the terms $\mathcal{P}_{\alpha}$ in the $N=8$ mean-field partition function at room temperature ($\sim$ 293 K). Starting from the left panel, the pressures plotted are 0, 4, 8, and 12 GPa. Blue points refer to experimentally realized polymorphs while red points refer to potential polymorphs hitherto not synthesized.}
\end{figure*}
While the analysis of Sec. \ref{sec:stat_mech} postulates a correlation between features of the single particle partition function in Eq. \ref{eq:real_one} and the tendency of ions in a crystal to join particular structures, a full accounting of the statistics of nucleation and growth would involve calculations of interfacial and/or surface energies \cite{Khaliullin2011}. Alternatively, given that nucleation of crystals in silicon is thought to occur on the order of a few nanometers, or roughly a few dozen atoms \cite{:/content/aip/journal/apl/79/10/10.1063/1.1401089}, we reason that features in the PES, which display scale invariance with respect to unit cell size, will be more likely to persist at the nucleation scale. For this reason, we calculated probabilities of occurrence over unit cell sizes $N = 8, 12, 16,$ and $24$. This choice of sampling allows for the formation of polymorphs with primitive unit cell size commensurate to any of those discussed in Sec. \ref{sec:exp_real}. However, due to the exponential growth of the number of local minima, which enter into the PES as a function of system size \cite{PhysRevE.59.48, stillinger2015energy}, and hence the large number of relaxations required to reach convergence of the $\mathcal{P}_{\alpha}$, $N=24$ represents a current rough upper limit on the unit cell size for which good statistics can be generated for silicon.

\begin{figure*}
\includegraphics[width=\linewidth]{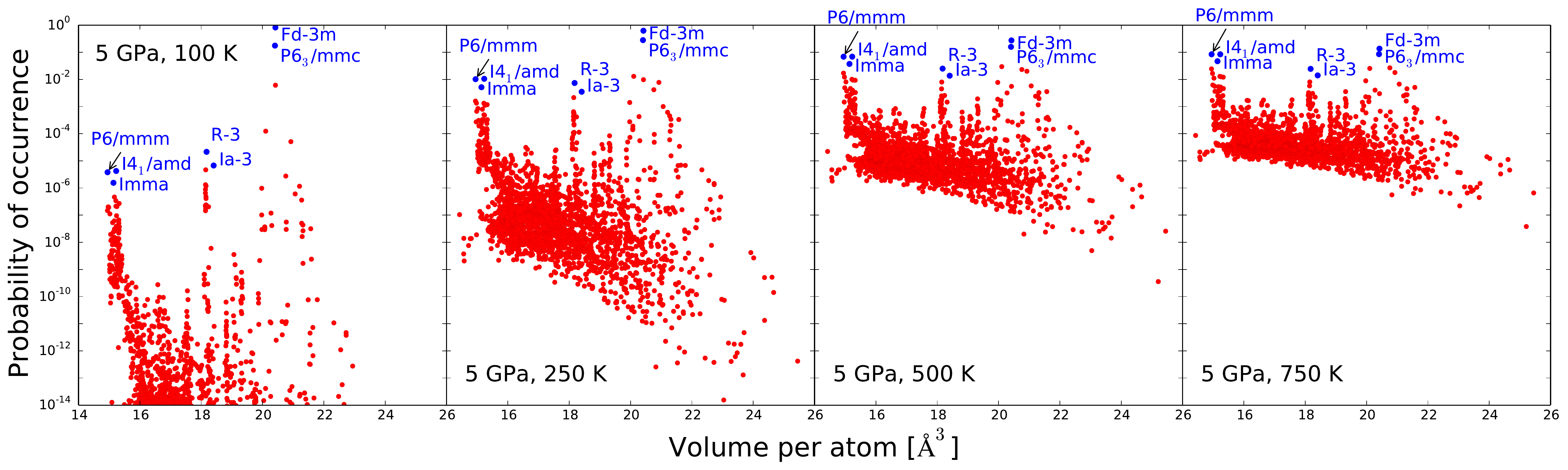}
\caption{\label{fig:temp_deps} Plot of the terms $\mathcal{P}_{\alpha}$ in the $N=8$ mean-field partition function at a fixed pressure of 5 GPa. Starting from the left panel, the temperatures plotted are 100, 250, 500, and 750 K. Blue points refer to experimentally realized polymorphs while red points refer to potential polymorphs hitherto not synthesized.}
\end{figure*}

\begin{figure*}
\includegraphics[width=\linewidth]{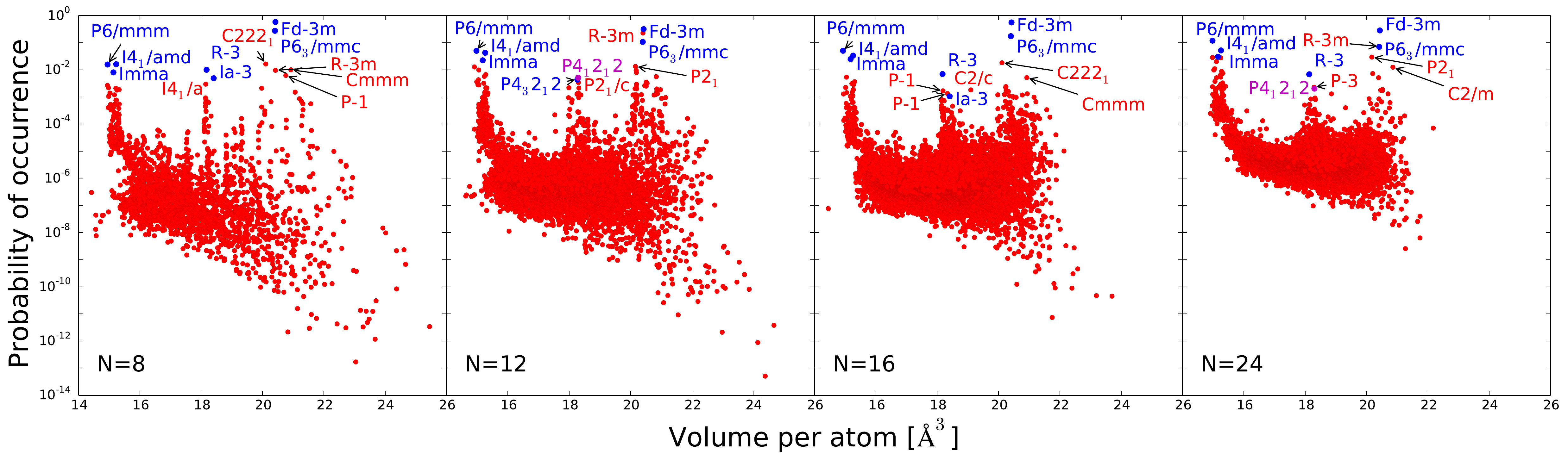}
\caption{\label{fig:N_deps} Plot of the terms $\mathcal{P}_{\alpha}$ in the mean-field partition function at a fixed pressure of 5 GPa and room temperature ($\sim$ 293 K). Starting from the left panel, the unit cell sizes plotted are N=8, 12, 16,  and 24. Blue points refer to experimentally realized polymorphs while red points refer to potential polymorphs hitherto not synthesized. The magenta point denotes ambiguity regarding synthesis.}
\end{figure*}

A total of $5,845$ random structures were relaxed with a unit cell size of N=8, which produced $2,597$ structural equivalence classes, each corresponding to a unique structure. The choice of N=8 as our initial unit cell size to consider stems from the fact that the primitive unit cell sizes of the Fd$\bar{3}$m, P6$_3$/mmc, R$\bar{3}$, and Ia$\bar{3}$ phases are N=2, 4, 8, and 8 respectively, with N=8 being the least common multiple of those known primitive unit cell sizes. Note that N=8 also accommodates I4$_1$/amd, Imma, and P6/mmm with primitive unit cells of N=2, 2, and 1 respectively. Fig. \ref{fig:press_deps} shows the dependence of the N=8 probability distribution on pressure plotted against the natural order parameter for pressure- and temperature-induced phase transitions: volume per atom. First, it is interesting to note that of the volume range accessed, not all structural volumes are equally probable. A common feature to all panels in Fig. \ref{fig:press_deps} (and also Figs. \ref{fig:temp_deps} and \ref{fig:N_deps}), there are regions of comparatively high probability, which occur at roughly 15 {\AA}$^{3}$ per atom, from 18 to 18.5 {\AA}$^{3}$ per atom, and again at around 20.5 {\AA}$^{3}$ per atom. We will call these regions ``jets'', and in contrast to the regions at around 15.5 - 18 {\AA}$^{3}$ per atom, 18.5 - 20 {\AA}$^{3}$ per atom, and greater than about 21 {\AA}$^{3}$ per atom (which we term ``hulls''), these high probabilities contribute the most strongly to the partition function. Therefore, it is significant that the most probably occurring structures, which cap each ``jet'' in Fig. \ref{fig:press_deps} are experimentally realized polymorphs as will be discussed more extensively below. Looking forward, discovery of these features in a new compound's PES would point towards where in parameter space various high-probability structures might likely become accessible to experiment.

At $0$ GPa (left-most panel of Fig. \ref{fig:press_deps}), the two clearly resonant probabilities are those at the tip of the 20.5 {\AA}$^{3}$ per atom jet, the experimentally realized Fd-$\bar{3}$m and P6$_3$/mmc phases. By contrast, the next most probable structure is an order of magnitude less probable, and the two other resonant jets at $15$ and 18-18.5 {\AA}$^{3}$ per atom fall well below this. At 12 GPa (right-most panel), the $15$ {\AA}$^{3}/$atom jet is clearly the most prevalent, capped by I4$_1$/amd, Imma, and P6/mmm, which are the first three experimental phases found upon static compression in a DAC. The intermediate pressures, $4$ and $8$ GPa (center-left and center-right panels), show situations where all three jets compete probabilistically with 8 GPa being an approximate coexistence point. That is, at 8 GPa the experimentally realized metastable R$\bar{3}$ and Ia$\bar{3}$ polymorphs, which are the two most probable structures around 18-18.5 {\AA}$^{3}$ per atom nearly approach P6$_3$/mmc in probability. We note that each of these structures (blue points in Fig. \ref{fig:press_deps}) was verified against its corresponding entry in the Inorganic Crystal Structure Database (ICSD) \cite{doi:10.1080/08893110410001664882}. Upon identifying high probability polymorphs, structures were brought into a standard primitive cell form using AFLOW online \cite{Curtarolo2012218}, and subsequently checked for dynamical stability by running phonon calculations with Quantum Espresso \cite{QE-2009}. Fd$\bar{3}$m, P6$_3$/mmc, Ia$\bar{3}$, and I4$_1$/amd were all determined to be stable, while R$\bar{3}$, Imma and P6/mmm were not. This is consistent with the experimental observations that P6/mmm and Imma transform reversibly under slow pressure release back to I4$_1$/amd and that R$\bar{3}$ typically only forms as a small fraction of a mixed phase with Ia$\bar{3}$. Interestingly, the latter fact seems to suggest that ions find their way into the Ia$\bar{3}$ structure by first identifying R$\bar{3}$ and then further proceeding to Ia$\bar{3}$.

The pressure dependence of our model gives us a good first-pass estimate of which structures will be able to compete thermodynamically under various pressurized synthesis conditions. In a similar manner, the temperature dependence of the probability distribution shown in Fig. \ref{fig:temp_deps} shows qualitatively how certain states can become more competitive with increasing temperature. It demonstrates a vertical squeeze in the distribution at elevated temperatures, which results in less probable structures becoming more competitive with more probable structures. The real upshot of both Figs. \ref{fig:press_deps} and \ref{fig:temp_deps} however, is that the ordering principle for experimental realizability is rather insensitive to model parameters. So long as jets and hulls can be identified in the probability distribution, then experimental realizability can be inferred from which structures are most probable within each jet. From those results, one can then proceed to assess dynamical stability, and calculate transition pressures and kinetic barriers.

In order to check for unit cell size dependence of the probability distribution, the $\mathcal{P}_{\alpha}$ for N=8 are again plotted at 5 GPa and 293 K in the left panel of Fig. \ref{fig:N_deps} against volume per atom. 5 GPa and 293 K are chosen so as to be able to easily compare distribution features across the various unit cell sizes without biasing the distributions towards any one volume regime. With the three jets discernible at around 15, 18-18.5, and 20.5 {\AA}$^{3}$ per atom, the Fd$\bar{3}$m and P6$_3$/mmc phases are the most probable at large volumes, R$\bar{3}$ and Ia$\bar{3}$ the most probable at intermediate volumes, and I4$_1$/amd, Imma, and the simple hexagonal phase the most probable at small volumes. Among the structures we found at N=8 with probabilities lower than those, which have been experimentally realized, we were able to identify the C222$_1$ structure predicted by Botti et al. to be a quasi-direct gap semiconductor suitable for photovoltaic applications \cite{PhysRevB.86.121204}. And while we do find a family of dissimilar I4$_1$/a structures beneath R$\bar{3}$ and Ia$\bar{3}$ at around $18$ {\AA}$^3$/atom, none of the more probably occurring structures could be identified with the (bt8, I4$_1$/a) structure synthesized by Rapp et al. \cite{rapp2015experimental} This is unsurprising since the development of isothermal-isobaric statistics takes thermodynamic equilibrium as an axiom, while ultrafast laser-induced confined microexplosion is a decidedly non-equilibrium technique.

Upon doubling the unit cell to N=16, 14,813 random structures were relaxed, generating $12,854$ structural equivalence classes. The resulting probability distribution is plotted in the center-right panel of Fig. \ref{fig:N_deps}. The only major discrepancy among the most probably occurring structures in each volume regime is that Ia$\bar{3}$ falls below a pair of P-1 structures. The P-1 structure with smaller volume per atom and larger probability of occurrence was found to be dynamically stable while the one with larger volume per atom and lower probability of occurrence was found to be dynamically unstable. Calculation of kinetic barriers is beyond the scope of this paper, but it could be the case that the dynamically stable P-1 structure is in kinetically unstable, explaining its absence in nature. A similar argument might be applied to the C2/c outlier found in the larger-volume hull of the center-right panel of Fig. \ref{fig:N_deps}. In any case, since R$\bar{3}$ constitutes a simple distortion of Ia$\bar{3}$ \cite{doi:10.1063/1.4962984}, it is likely appropriate to consider the two polymorphs as belonging to the same funnel of attraction. Such a renormalization of the Ia$\bar{3}$ probability would render it as jointly the most probably occurring structure around the $\sim$ 18-18.5 {\AA}$^3$ per atom jet and more probably occurring than the C2/c structure.

For good measure, we generated probabilities of occurrence for N=12, since at least one polymorph has been both theoretically predicted and experimentally synthesized by non-equilibrium methods with a 12 atom primitive unit cell \cite{rapp2015experimental}. For N=12, 9,919 random structures were relaxed, which resulted in $6,655$ equivalence classes. Note that while Fd-$\bar{3}$m, P6$_3$/mmc, I4$_1$/amd, Imma, and P6/mmm are compatible with N=12 since their primitive unit cell sizes divide 12 evenly, R$\bar{3}$ and Ia$\bar{3}$ are incompatible with N=12 since they both have a primitive unit cell of N=8. As such, we should not expect the appearance of either R$\bar{3}$ or Ia$\bar{3}$ at N=12. At small volumes,  I4$_1$/amd, Imma, and P6/mmm still squarely dominate the probability distribution. Meanwhile, the top two most probably occurring polymorphs at moderate volumes have space groups P4$_1$2$_1$2 and P4$_3$2$_1$2 followed by a family of inequivalent P2$_1$/c structures. Phonon calculations were carried out for P4$_1$2$_1$2 and P4$_3$2$_1$2, and both are confirmed to be dynamically stable. Surprisingly, the P4$_3$2$_1$2 structure is exactly the phase recently synthesized under distinctly far-from-equilibrium conditions by Rapp et al. \cite{rapp2015experimental} (st12, P4$_3$2$_1$2). This stands in contrast to the same authors' I4$_1$/a structure, which does not appear with high probability at N=8. More generally, other than P4$_3$2$_1$2 we do not recover as probably occurring any of the other polymorphs, which have successfully been synthesized under non-equilibrium or exclusively chemically altered conditions. P4$_1$2$_1$ (Si-VIII) and P4$_2$2 (Si-IX) are included in this classification since they are formed via rapid pressure release, and as regards chemical precursor synthesis routes, one would need to explore the PES of the modified chemistry rather than that of pure silicon in order to derive accurate statistics. On the one hand, we would not necessarily expect an equilibrium model to capture non-equilibrium phenomena (although it is not expressly forbidden). But in the case of P4$_3$2$_1$2, it could be that the frequency of occurrence for this state drives formation even when the Boltzmann factor becomes inapplicable due to the presence of the microscopic temperature, pressure, and electric field inhomogeneities likely introduced through ultrafast laser-induced confined microexplosion. Alternatively, there could be some semblance of local thermodynamic equilibrium even under these extreme conditions such that the Boltzmann factor still contributes locally to structure formation.

Identification of the P4$_1$2$_1$2 phase is less straight-forward, but it might be associated with the P4$_1$2$_1$2 (Si-XIII) phase formed experimentally by annealing an indentation loading sample of mixed R$\bar{3}$/Ia$\bar{3}$ at 473 K \cite{doi:10.1063/1.3124366}. Multiple predictions have previously been put forward to solve the crystal structure of this phase \cite{mujica2015low,doi:10.1021/ja304380p}, but its obstinacy towards coexistence with other phases has made any sort of definite characterization challenging. It is noteworthy that the experimental P4$_1$2$_1$2 phase only occurs upon annealing the particular mixture of R$\bar{3}$/Ia$\bar{3}$ produced by indentation loading, and has not to this point been obtained as a phase pure isolate. At large volumes, a new structure with space group R$\bar{3}$m appears between the Fd$\bar{3}$m and P6$_3$/mmc phases with 12 atoms in its primitive unit cell (inequivalent to the R$\bar{3}$m structure found at N=8). Phonon calculations indicate that this phase is also dynamically stable, which would seem to suggest its viability as an experimentally realizable silicon polymorph. However, upon doubling the unit cell size from 12 to 24, a drop in the R$\bar{3}$m probability occurs with respect to the P6$_3$/mmc phase (the plot points for the two phases are overlaid). Extrapolating that relative drop to larger unit cell sizes would see a continued decrease in the probability of R$\bar{3}$m and thus a restoration of P6$_3$/mmc as the second most probably occurring at large volumes.
 
 It should be noted that for N=24, the relative complexity and expansiveness of the PES prohibited us computationally from fully exploring funnel of attraction statistics, but we were able to extract some features of it from the 9,845 successfully relaxed random structures, sorted into 9,166 equivalence classes. The prevalence of I4$_1$/amd, Imma, and P6/mmm at small volumes is replicated at N=24, while R$\bar{3}$ remains the most probable intermediate-volume structure. As can be seen in the right-most panel of Fig. \ref{fig:N_deps}, we were unable to encounter the Ia$\bar{3}$ phase among any of the successfully relaxed and sorted structures. However, the argument that Ia$\bar{3}$ and R$\bar{3}$ share a common funnel of attraction would again place Ia$\bar{3}$ alongside R$\bar{3}$ as the most probable intermediate-volume structure. Below the R$\bar{3}$/Ia$\bar{3}$ probability, we find a structure with space group P$\bar{3}$ only slightly above the 12 atom P4$_1$2$_1$2 structure. This relative drop in probability could indicate either a reason for why phase-pure Si-XIII is difficult to achieve or alternatively, that our P4$_1$2$_1$2 structure is not assignable to Si-XIII and is some entirely different phase, which would be difficult to synthesize. Without further experimental input such as X-ray diffraction data on a phase-pure sample, it is challenging to determine which scenario reflects reality. Also missing from the N=24 probability distribution as being high-probability is the P4$_3$2$_1$2 structure, potentially explaining why R$\bar{3}$/Ia$\bar{3}$ are easier to synthesize under equilibrium conditions, and why non-equilibrium methods are needed to realize P4$_3$2$_1$2. And conspicuously missing from any unit cell size are any phases with volume below about 14 {\AA}$^{3}$ per atom. \footnote{This observation might suggest that our calculations overestimate the volume of P6/mmm, which is typically cited to be between 13 to 14 {\AA}$^{3}$ per atom. However, that is the equilibrium volume under pressure, which will generally be smaller than the volume at ambient conditions} Figs. \ref{fig:press_deps}, \ref{fig:temp_deps}, and \ref{fig:N_deps} all show that no such structures were found with our sampling technique. One explanation for this is that at pressures far above 15 GPa (roughly where P6/mmm is accessed), the potential enthalpy surface is so dissimilar to the zero pressure PES that high pressure local minima cease to be minima at all at ambient conditions.

There are three main approximations we have performed in order to be able to relate ab initio exploration of silicon's PES statistics to the experimental formation of its polymorphs. First, we have chosen to use a mean field, single particle model to represent the statistical mechanics of crystal formation in a real solid, a process that in reality is dynamical and proceeds by nucleation and growth. Second, we have replaced all intra-basin contributions to the configurational integral in Eq. \ref{eq:my_seven} by those evaluated at basin minima, which corresponds to flattening each basin into a square well. And finally, we have approximated the true potential enthalpy per particle by $\psi^{approx.}$, calculated from potential energy and volume data generated at zero temperature and pressure. Despite these approximations, the qualitative trends in Fig. \ref{fig:N_deps} are clear and indicate that of the thousands of potential polymorphs found, Fd$\bar{3}$m, P6$_3$/mmc, R$\bar{3}$/Ia$\bar{3}$, I4$_1$/amd, Imma, and P6/mmm  are the most persistently probable in their respective volume regimes and thus will be the most likely to be accessed under equilibrium conditions, corroborating the experimental reality. Of these, R$\bar{3}$, Ia$\bar{3}$, and P6$_3$/mmc are known to be metastable. Thus, we have accurately reproduced the equilibrium-formed, metastable sector of silicon's polymorph spectrum using quantities readily produced by ab initio techniques (total energies and volumes) by gathering statistics on random structure relaxations, variants of which are also currently employed by the structure prediction community \cite{0953-8984-23-5-053201}. Our model therefore provides a tool by which candidate polymorphs of little-studied compounds can be easily assessed for experimental realization in the future.
%
\section{\label{sec:conclusion}Conclusion}
%
In addition to their novel properties, metastable polymorphs such as the diamond phase of carbon see wide industrial application because of their amenability to being artificially synthesized in bulk quantities \cite{USGS}. Given that equilibrium materials synthesis techniques have historically been more extensively studied and hence, have been used to greater effect to synthesize materials at the industrial scale, it is important to be able to understand which of the many candidate polymorphs produced by an ab initio search are likely to be synthesized under equilibrium conditions. We have therefore presented a mean-field model of experimental polymorph realizability predicated on the isothermal-isobaric statistics of a single, interacting ion. Using this formalism, we demonstrated that the polymorphs of silicon synthesizable under near-equilibrium conditions can be separated from the many other candidate polymorphs produced by an ab initio search by selection of resonant probabilities in the resulting distribution. By way of interpretation, we have shown that in addition to global enthalpic favorability, ions are driven into specific configurations by a measure of entropy, which is related to the hypervolume of a structure's funnel of attraction. This approach provides a prescription whereby candidate polymorphs uncovered for compounds, which have previously seen little experimental work, can be ranked for experimental consideration by their probabilities of occurrence.


\begin{acknowledgments}
This work was supported as part of the Center for the Next Generation of Materials by Design, an Energy Frontier Research Center funded by the U.S. Department of Energy, Office of Science, Basic Energy Sciences. The research was performed using computational resources sponsored by the Department of Energy's Office of Energy Efficiency and Renewable Energy and located at the National Renewable Energy Laboratory.
\end{acknowledgments}


\begin{thebibliography}{35}%
\makeatletter
\providecommand \@ifxundefined [1]{%
 \@ifx{#1\undefined}
}%
\providecommand \@ifnum [1]{%
 \ifnum #1\expandafter \@firstoftwo
 \else \expandafter \@secondoftwo
 \fi
}%
\providecommand \@ifx [1]{%
 \ifx #1\expandafter \@firstoftwo
 \else \expandafter \@secondoftwo
 \fi
}%
\providecommand \natexlab [1]{#1}%
\providecommand \enquote  [1]{``#1''}%
\providecommand \bibnamefont  [1]{#1}%
\providecommand \bibfnamefont [1]{#1}%
\providecommand \citenamefont [1]{#1}%
\providecommand \href@noop [0]{\@secondoftwo}%
\providecommand \href [0]{\begingroup \@sanitize@url \@href}%
\providecommand \@href[1]{\@@startlink{#1}\@@href}%
\providecommand \@@href[1]{\endgroup#1\@@endlink}%
\providecommand \@sanitize@url [0]{\catcode `\\12\catcode `\$12\catcode
  `\&12\catcode `\#12\catcode `\^12\catcode `\_12\catcode `\%12\relax}%
\providecommand \@@startlink[1]{}%
\providecommand \@@endlink[0]{}%
\providecommand \url  [0]{\begingroup\@sanitize@url \@url }%
\providecommand \@url [1]{\endgroup\@href {#1}{\urlprefix }}%
\providecommand \urlprefix  [0]{URL }%
\providecommand \Eprint [0]{\href }%
\providecommand \doibase [0]{http://dx.doi.org/}%
\providecommand \selectlanguage [0]{\@gobble}%
\providecommand \bibinfo  [0]{\@secondoftwo}%
\providecommand \bibfield  [0]{\@secondoftwo}%
\providecommand \translation [1]{[#1]}%
\providecommand \BibitemOpen [0]{}%
\providecommand \bibitemStop [0]{}%
\providecommand \bibitemNoStop [0]{.\EOS\space}%
\providecommand \EOS [0]{\spacefactor3000\relax}%
\providecommand \BibitemShut  [1]{\csname bibitem#1\endcsname}%
\let\auto@bib@innerbib\@empty
\bibitem [{\citenamefont {Haberl}\ \emph {et~al.}(2016)\citenamefont {Haberl},
  \citenamefont {Strobel},\ and\ \citenamefont
  {Bradby}}]{doi:10.1063/1.4962984}%
  \BibitemOpen
  \bibfield  {author} {\bibinfo {author} {\bibfnamefont {B.}~\bibnamefont
  {Haberl}}, \bibinfo {author} {\bibfnamefont {T.~A.}\ \bibnamefont {Strobel}},
  \ and\ \bibinfo {author} {\bibfnamefont {J.~E.}\ \bibnamefont {Bradby}},\
  }\href {\doibase 10.1063/1.4962984} {\bibfield  {journal} {\bibinfo
  {journal} {Applied Physics Reviews}\ }\textbf {\bibinfo {volume} {3}},\
  \bibinfo {pages} {040808} (\bibinfo {year} {2016})},\ \Eprint
  {http://arxiv.org/abs/http://dx.doi.org/10.1063/1.4962984}
  {http://dx.doi.org/10.1063/1.4962984} \BibitemShut {NoStop}%
\bibitem [{\citenamefont {Jamieson}(1963)}]{Jamieson762}%
  \BibitemOpen
  \bibfield  {author} {\bibinfo {author} {\bibfnamefont {J.~C.}\ \bibnamefont
  {Jamieson}},\ }\href {\doibase 10.1126/science.139.3556.762} {\bibfield
  {journal} {\bibinfo  {journal} {Science}\ }\textbf {\bibinfo {volume}
  {139}},\ \bibinfo {pages} {762} (\bibinfo {year} {1963})},\ \Eprint
  {http://arxiv.org/abs/http://science.sciencemag.org/content/139/3556/762.full.pdf}
  {http://science.sciencemag.org/content/139/3556/762.full.pdf} \BibitemShut
  {NoStop}%
\bibitem [{\citenamefont {Rapp}\ \emph {et~al.}(2015)\citenamefont {Rapp},
  \citenamefont {Haberl}, \citenamefont {Pickard}, \citenamefont {Bradby},
  \citenamefont {Gamaly}, \citenamefont {Williams},\ and\ \citenamefont
  {Rode}}]{rapp2015experimental}%
  \BibitemOpen
  \bibfield  {author} {\bibinfo {author} {\bibfnamefont {L.}~\bibnamefont
  {Rapp}}, \bibinfo {author} {\bibfnamefont {B.}~\bibnamefont {Haberl}},
  \bibinfo {author} {\bibfnamefont {C.}~\bibnamefont {Pickard}}, \bibinfo
  {author} {\bibfnamefont {J.}~\bibnamefont {Bradby}}, \bibinfo {author}
  {\bibfnamefont {E.~G.}\ \bibnamefont {Gamaly}}, \bibinfo {author}
  {\bibfnamefont {J.}~\bibnamefont {Williams}}, \ and\ \bibinfo {author}
  {\bibfnamefont {A.~V.}\ \bibnamefont {Rode}},\ }\href@noop {} {\bibfield
  {journal} {\bibinfo  {journal} {Nature communications}\ }\textbf {\bibinfo
  {volume} {6}} (\bibinfo {year} {2015})}\BibitemShut {NoStop}%
\bibitem [{\citenamefont {Kim}\ \emph {et~al.}(2015)\citenamefont {Kim},
  \citenamefont {Stefanoski}, \citenamefont {Kurakevych},\ and\ \citenamefont
  {Strobel}}]{Kim2015}%
  \BibitemOpen
  \bibfield  {author} {\bibinfo {author} {\bibfnamefont {D.~Y.}\ \bibnamefont
  {Kim}}, \bibinfo {author} {\bibfnamefont {S.}~\bibnamefont {Stefanoski}},
  \bibinfo {author} {\bibfnamefont {O.~O.}\ \bibnamefont {Kurakevych}}, \ and\
  \bibinfo {author} {\bibfnamefont {T.~A.}\ \bibnamefont {Strobel}},\ }\href
  {http://dx.doi.org/10.1038/nmat4140} {\bibfield  {journal} {\bibinfo
  {journal} {Nat Mater}\ }\textbf {\bibinfo {volume} {14}},\ \bibinfo {pages}
  {169} (\bibinfo {year} {2015})},\ \bibinfo {note} {letter}\BibitemShut
  {NoStop}%
\bibitem [{\citenamefont {McMahon}\ and\ \citenamefont
  {Nelmes}(1993)}]{PhysRevB.47.8337}%
  \BibitemOpen
  \bibfield  {author} {\bibinfo {author} {\bibfnamefont {M.~I.}\ \bibnamefont
  {McMahon}}\ and\ \bibinfo {author} {\bibfnamefont {R.~J.}\ \bibnamefont
  {Nelmes}},\ }\href {\doibase 10.1103/PhysRevB.47.8337} {\bibfield  {journal}
  {\bibinfo  {journal} {Phys. Rev. B}\ }\textbf {\bibinfo {volume} {47}},\
  \bibinfo {pages} {8337} (\bibinfo {year} {1993})}\BibitemShut {NoStop}%
\bibitem [{\citenamefont {McMahon}\ \emph {et~al.}(1994)\citenamefont
  {McMahon}, \citenamefont {Nelmes}, \citenamefont {Wright},\ and\
  \citenamefont {Allan}}]{PhysRevB.50.739}%
  \BibitemOpen
  \bibfield  {author} {\bibinfo {author} {\bibfnamefont {M.~I.}\ \bibnamefont
  {McMahon}}, \bibinfo {author} {\bibfnamefont {R.~J.}\ \bibnamefont {Nelmes}},
  \bibinfo {author} {\bibfnamefont {N.~G.}\ \bibnamefont {Wright}}, \ and\
  \bibinfo {author} {\bibfnamefont {D.~R.}\ \bibnamefont {Allan}},\ }\href
  {\doibase 10.1103/PhysRevB.50.739} {\bibfield  {journal} {\bibinfo  {journal}
  {Phys. Rev. B}\ }\textbf {\bibinfo {volume} {50}},\ \bibinfo {pages} {739}
  (\bibinfo {year} {1994})}\BibitemShut {NoStop}%
\bibitem [{\citenamefont {Olijnyk}\ \emph {et~al.}(1984)\citenamefont
  {Olijnyk}, \citenamefont {Sikka},\ and\ \citenamefont
  {Holzapfel}}]{OLIJNYK1984137}%
  \BibitemOpen
  \bibfield  {author} {\bibinfo {author} {\bibfnamefont {H.}~\bibnamefont
  {Olijnyk}}, \bibinfo {author} {\bibfnamefont {S.}~\bibnamefont {Sikka}}, \
  and\ \bibinfo {author} {\bibfnamefont {W.}~\bibnamefont {Holzapfel}},\ }\href
  {\doibase http://dx.doi.org/10.1016/0375-9601(84)90219-6} {\bibfield
  {journal} {\bibinfo  {journal} {Physics Letters A}\ }\textbf {\bibinfo
  {volume} {103}},\ \bibinfo {pages} {137 } (\bibinfo {year}
  {1984})}\BibitemShut {NoStop}%
\bibitem [{\citenamefont {Minomura}\ and\ \citenamefont
  {Drickamer}(1962)}]{MINOMURA1962451}%
  \BibitemOpen
  \bibfield  {author} {\bibinfo {author} {\bibfnamefont {S.}~\bibnamefont
  {Minomura}}\ and\ \bibinfo {author} {\bibfnamefont {H.}~\bibnamefont
  {Drickamer}},\ }\href {\doibase
  http://dx.doi.org/10.1016/0022-3697(62)90085-9} {\bibfield  {journal}
  {\bibinfo  {journal} {Journal of Physics and Chemistry of Solids}\ }\textbf
  {\bibinfo {volume} {23}},\ \bibinfo {pages} {451 } (\bibinfo {year}
  {1962})}\BibitemShut {NoStop}%
\bibitem [{\citenamefont {Wentorf}\ and\ \citenamefont
  {Kasper}(1963)}]{Wentorf338}%
  \BibitemOpen
  \bibfield  {author} {\bibinfo {author} {\bibfnamefont {R.~H.}\ \bibnamefont
  {Wentorf}}\ and\ \bibinfo {author} {\bibfnamefont {J.~S.}\ \bibnamefont
  {Kasper}},\ }\href {\doibase 10.1126/science.139.3552.338-a} {\bibfield
  {journal} {\bibinfo  {journal} {Science}\ }\textbf {\bibinfo {volume}
  {139}},\ \bibinfo {pages} {338} (\bibinfo {year} {1963})},\ \Eprint
  {http://arxiv.org/abs/http://science.sciencemag.org/content/139/3552/338.2.full.pdf}
  {http://science.sciencemag.org/content/139/3552/338.2.full.pdf} \BibitemShut
  {NoStop}%
\bibitem [{\citenamefont {Ruffell}\ \emph {et~al.}(2011)\citenamefont
  {Ruffell}, \citenamefont {Sears}, \citenamefont {Knights}, \citenamefont
  {Bradby},\ and\ \citenamefont {Williams}}]{PhysRevB.83.075316}%
  \BibitemOpen
  \bibfield  {author} {\bibinfo {author} {\bibfnamefont {S.}~\bibnamefont
  {Ruffell}}, \bibinfo {author} {\bibfnamefont {K.}~\bibnamefont {Sears}},
  \bibinfo {author} {\bibfnamefont {A.~P.}\ \bibnamefont {Knights}}, \bibinfo
  {author} {\bibfnamefont {J.~E.}\ \bibnamefont {Bradby}}, \ and\ \bibinfo
  {author} {\bibfnamefont {J.~S.}\ \bibnamefont {Williams}},\ }\href {\doibase
  10.1103/PhysRevB.83.075316} {\bibfield  {journal} {\bibinfo  {journal} {Phys.
  Rev. B}\ }\textbf {\bibinfo {volume} {83}},\ \bibinfo {pages} {075316}
  (\bibinfo {year} {2011})}\BibitemShut {NoStop}%
\bibitem [{\citenamefont {Weill}\ \emph {et~al.}(1989)\citenamefont {Weill},
  \citenamefont {Mansot}, \citenamefont {Sagon}, \citenamefont {Carlone},\ and\
  \citenamefont {Besson}}]{0268-1242-4-4-029}%
  \BibitemOpen
  \bibfield  {author} {\bibinfo {author} {\bibfnamefont {G.}~\bibnamefont
  {Weill}}, \bibinfo {author} {\bibfnamefont {J.~L.}\ \bibnamefont {Mansot}},
  \bibinfo {author} {\bibfnamefont {G.}~\bibnamefont {Sagon}}, \bibinfo
  {author} {\bibfnamefont {C.}~\bibnamefont {Carlone}}, \ and\ \bibinfo
  {author} {\bibfnamefont {J.~M.}\ \bibnamefont {Besson}},\ }\href
  {http://stacks.iop.org/0268-1242/4/i=4/a=029} {\bibfield  {journal} {\bibinfo
   {journal} {Semiconductor Science and Technology}\ }\textbf {\bibinfo
  {volume} {4}},\ \bibinfo {pages} {280} (\bibinfo {year} {1989})}\BibitemShut
  {NoStop}%
\bibitem [{\citenamefont {Ruffell}\ \emph {et~al.}(2009)\citenamefont
  {Ruffell}, \citenamefont {Haberl}, \citenamefont {Koenig}, \citenamefont
  {Bradby},\ and\ \citenamefont {Williams}}]{doi:10.1063/1.3124366}%
  \BibitemOpen
  \bibfield  {author} {\bibinfo {author} {\bibfnamefont {S.}~\bibnamefont
  {Ruffell}}, \bibinfo {author} {\bibfnamefont {B.}~\bibnamefont {Haberl}},
  \bibinfo {author} {\bibfnamefont {S.}~\bibnamefont {Koenig}}, \bibinfo
  {author} {\bibfnamefont {J.~E.}\ \bibnamefont {Bradby}}, \ and\ \bibinfo
  {author} {\bibfnamefont {J.~S.}\ \bibnamefont {Williams}},\ }\href {\doibase
  10.1063/1.3124366} {\bibfield  {journal} {\bibinfo  {journal} {Journal of
  Applied Physics}\ }\textbf {\bibinfo {volume} {105}},\ \bibinfo {pages}
  {093513} (\bibinfo {year} {2009})},\ \Eprint
  {http://arxiv.org/abs/http://dx.doi.org/10.1063/1.3124366}
  {http://dx.doi.org/10.1063/1.3124366} \BibitemShut {NoStop}%
\bibitem [{\citenamefont {Hanfland}\ \emph {et~al.}(1999)\citenamefont
  {Hanfland}, \citenamefont {Schwarz}, \citenamefont {Syassen},\ and\
  \citenamefont {Takemura}}]{PhysRevLett.82.1197}%
  \BibitemOpen
  \bibfield  {author} {\bibinfo {author} {\bibfnamefont {M.}~\bibnamefont
  {Hanfland}}, \bibinfo {author} {\bibfnamefont {U.}~\bibnamefont {Schwarz}},
  \bibinfo {author} {\bibfnamefont {K.}~\bibnamefont {Syassen}}, \ and\
  \bibinfo {author} {\bibfnamefont {K.}~\bibnamefont {Takemura}},\ }\href
  {\doibase 10.1103/PhysRevLett.82.1197} {\bibfield  {journal} {\bibinfo
  {journal} {Phys. Rev. Lett.}\ }\textbf {\bibinfo {volume} {82}},\ \bibinfo
  {pages} {1197} (\bibinfo {year} {1999})}\BibitemShut {NoStop}%
\bibitem [{\citenamefont {Duclos}\ \emph {et~al.}(1990)\citenamefont {Duclos},
  \citenamefont {Vohra},\ and\ \citenamefont {Ruoff}}]{PhysRevB.41.12021}%
  \BibitemOpen
  \bibfield  {author} {\bibinfo {author} {\bibfnamefont {S.~J.}\ \bibnamefont
  {Duclos}}, \bibinfo {author} {\bibfnamefont {Y.~K.}\ \bibnamefont {Vohra}}, \
  and\ \bibinfo {author} {\bibfnamefont {A.~L.}\ \bibnamefont {Ruoff}},\ }\href
  {\doibase 10.1103/PhysRevB.41.12021} {\bibfield  {journal} {\bibinfo
  {journal} {Phys. Rev. B}\ }\textbf {\bibinfo {volume} {41}},\ \bibinfo
  {pages} {12021} (\bibinfo {year} {1990})}\BibitemShut {NoStop}%
\bibitem [{\citenamefont {Zhao}\ \emph {et~al.}(1986)\citenamefont {Zhao},
  \citenamefont {Buehler}, \citenamefont {Sites},\ and\ \citenamefont
  {Spain}}]{ZHAO1986679}%
  \BibitemOpen
  \bibfield  {author} {\bibinfo {author} {\bibfnamefont {Y.-X.}\ \bibnamefont
  {Zhao}}, \bibinfo {author} {\bibfnamefont {F.}~\bibnamefont {Buehler}},
  \bibinfo {author} {\bibfnamefont {J.~R.}\ \bibnamefont {Sites}}, \ and\
  \bibinfo {author} {\bibfnamefont {I.~L.}\ \bibnamefont {Spain}},\ }\href
  {\doibase http://dx.doi.org/10.1016/0038-1098(86)90372-8} {\bibfield
  {journal} {\bibinfo  {journal} {Solid State Communications}\ }\textbf
  {\bibinfo {volume} {59}},\ \bibinfo {pages} {679 } (\bibinfo {year}
  {1986})}\BibitemShut {NoStop}%
\bibitem [{\citenamefont {Botti}\ \emph {et~al.}(2012)\citenamefont {Botti},
  \citenamefont {Flores-Livas}, \citenamefont {Amsler}, \citenamefont
  {Goedecker},\ and\ \citenamefont {Marques}}]{PhysRevB.86.121204}%
  \BibitemOpen
  \bibfield  {author} {\bibinfo {author} {\bibfnamefont {S.}~\bibnamefont
  {Botti}}, \bibinfo {author} {\bibfnamefont {J.~A.}\ \bibnamefont
  {Flores-Livas}}, \bibinfo {author} {\bibfnamefont {M.}~\bibnamefont
  {Amsler}}, \bibinfo {author} {\bibfnamefont {S.}~\bibnamefont {Goedecker}}, \
  and\ \bibinfo {author} {\bibfnamefont {M.~A.~L.}\ \bibnamefont {Marques}},\
  }\href {\doibase 10.1103/PhysRevB.86.121204} {\bibfield  {journal} {\bibinfo
  {journal} {Phys. Rev. B}\ }\textbf {\bibinfo {volume} {86}},\ \bibinfo
  {pages} {121204} (\bibinfo {year} {2012})}\BibitemShut {NoStop}%
\bibitem [{\citenamefont {Stillinger}(2015)}]{stillinger2015energy}%
  \BibitemOpen
  \bibfield  {author} {\bibinfo {author} {\bibfnamefont {F.~H.}\ \bibnamefont
  {Stillinger}},\ }\href {https://books.google.com/books?id=qKC4CgAAQBAJ}
  {\emph {\bibinfo {title} {Energy Landscapes, Inherent Structures, and
  Condensed-Matter Phenomena}}}\ (\bibinfo  {publisher} {Princeton University
  Press},\ \bibinfo {year} {2015})\BibitemShut {NoStop}%
\bibitem [{\citenamefont {Stevanovi\ifmmode~\acute{c}\else
  \'{c}\fi{}}(2016)}]{PhysRevLett.116.075503}%
  \BibitemOpen
  \bibfield  {author} {\bibinfo {author} {\bibfnamefont {V.}~\bibnamefont
  {Stevanovi\ifmmode~\acute{c}\else \'{c}\fi{}}},\ }\href {\doibase
  10.1103/PhysRevLett.116.075503} {\bibfield  {journal} {\bibinfo  {journal}
  {Phys. Rev. Lett.}\ }\textbf {\bibinfo {volume} {116}},\ \bibinfo {pages}
  {075503} (\bibinfo {year} {2016})}\BibitemShut {NoStop}%
\bibitem [{\citenamefont {Khaliullin}\ \emph {et~al.}(2011)\citenamefont
  {Khaliullin}, \citenamefont {Eshet}, \citenamefont {Kuhne}, \citenamefont
  {Behler},\ and\ \citenamefont {Parrinello}}]{Khaliullin2011}%
  \BibitemOpen
  \bibfield  {author} {\bibinfo {author} {\bibfnamefont {R.~Z.}\ \bibnamefont
  {Khaliullin}}, \bibinfo {author} {\bibfnamefont {H.}~\bibnamefont {Eshet}},
  \bibinfo {author} {\bibfnamefont {T.~D.}\ \bibnamefont {Kuhne}}, \bibinfo
  {author} {\bibfnamefont {J.}~\bibnamefont {Behler}}, \ and\ \bibinfo {author}
  {\bibfnamefont {M.}~\bibnamefont {Parrinello}},\ }\href {\doibase
  10.1038/nmat3078} {\bibfield  {journal} {\bibinfo  {journal} {Nat Mater}\
  }\textbf {\bibinfo {volume} {10}},\ \bibinfo {pages} {693} (\bibinfo {year}
  {2011})}\BibitemShut {NoStop}%
\bibitem [{\citenamefont {Verma}\ and\ \citenamefont
  {Krishna}(1966)}]{verma1966polymorphism}%
  \BibitemOpen
  \bibfield  {author} {\bibinfo {author} {\bibfnamefont {A.}~\bibnamefont
  {Verma}}\ and\ \bibinfo {author} {\bibfnamefont {P.}~\bibnamefont
  {Krishna}},\ }\href {https://books.google.com/books?id=oiqFAAAAIAAJ} {\emph
  {\bibinfo {title} {Polymorphism and polytypism in crystals}}},\ Wiley
  monographs in crystallography\ (\bibinfo  {publisher} {Wiley},\ \bibinfo
  {year} {1966})\BibitemShut {NoStop}%
\bibitem [{\citenamefont {Perdew}\ \emph {et~al.}(1996)\citenamefont {Perdew},
  \citenamefont {Burke},\ and\ \citenamefont
  {Ernzerhof}}]{PhysRevLett.77.3865}%
  \BibitemOpen
  \bibfield  {author} {\bibinfo {author} {\bibfnamefont {J.~P.}\ \bibnamefont
  {Perdew}}, \bibinfo {author} {\bibfnamefont {K.}~\bibnamefont {Burke}}, \
  and\ \bibinfo {author} {\bibfnamefont {M.}~\bibnamefont {Ernzerhof}},\ }\href
  {\doibase 10.1103/PhysRevLett.77.3865} {\bibfield  {journal} {\bibinfo
  {journal} {Phys. Rev. Lett.}\ }\textbf {\bibinfo {volume} {77}},\ \bibinfo
  {pages} {3865} (\bibinfo {year} {1996})}\BibitemShut {NoStop}%
\bibitem [{\citenamefont {Bl\"ochl}(1994)}]{PhysRevB.50.17953}%
  \BibitemOpen
  \bibfield  {author} {\bibinfo {author} {\bibfnamefont {P.~E.}\ \bibnamefont
  {Bl\"ochl}},\ }\href {\doibase 10.1103/PhysRevB.50.17953} {\bibfield
  {journal} {\bibinfo  {journal} {Phys. Rev. B}\ }\textbf {\bibinfo {volume}
  {50}},\ \bibinfo {pages} {17953} (\bibinfo {year} {1994})}\BibitemShut
  {NoStop}%
\bibitem [{\citenamefont {Kresse}\ and\ \citenamefont
  {Furthm?ller}(1996)}]{CMS.6.15}%
  \BibitemOpen
  \bibfield  {author} {\bibinfo {author} {\bibfnamefont {G.}~\bibnamefont
  {Kresse}}\ and\ \bibinfo {author} {\bibfnamefont {J.}~\bibnamefont
  {Furthm?ller}},\ }\href {\doibase 10.1016/0927-0256(96)00008-0} {\bibfield
  {journal} {\bibinfo  {journal} {Comput. Mater. Sci.}\ }\textbf {\bibinfo
  {volume} {6}},\ \bibinfo {pages} {15 } (\bibinfo {year} {1996})}\BibitemShut
  {NoStop}%
\bibitem [{\citenamefont {Press}\ \emph {et~al.}(2007)\citenamefont {Press},
  \citenamefont {Teukolsky}, \citenamefont {Vetterling},\ and\ \citenamefont
  {Flannery}}]{Press:2007:NRE:1403886}%
  \BibitemOpen
  \bibfield  {author} {\bibinfo {author} {\bibfnamefont {W.~H.}\ \bibnamefont
  {Press}}, \bibinfo {author} {\bibfnamefont {S.~A.}\ \bibnamefont
  {Teukolsky}}, \bibinfo {author} {\bibfnamefont {W.~T.}\ \bibnamefont
  {Vetterling}}, \ and\ \bibinfo {author} {\bibfnamefont {B.~P.}\ \bibnamefont
  {Flannery}},\ }\href@noop {} {\emph {\bibinfo {title} {Numerical Recipes 3rd
  Edition: The Art of Scientific Computing}}},\ \bibinfo {edition} {3rd}\ ed.\
  (\bibinfo  {publisher} {Cambridge University Press},\ \bibinfo {address} {New
  York, NY, USA},\ \bibinfo {year} {2007})\BibitemShut {NoStop}%
\bibitem [{pyl(2014)}]{pylada}%
  \BibitemOpen
  \href@noop {} {\enquote {\bibinfo {title} {{Pylada Computational
  Framework}},}\ }\bibinfo {howpublished}
  {\url{https://github.com/pylada/pylada-light}} (\bibinfo {year} {2014}),\
  \bibinfo {note} {accessed: 2016-12-08}\BibitemShut {NoStop}%
\bibitem [{\citenamefont {Sunkara}\ \emph {et~al.}(2001)\citenamefont
  {Sunkara}, \citenamefont {Sharma}, \citenamefont {Miranda}, \citenamefont
  {Lian},\ and\ \citenamefont
  {Dickey}}]{:/content/aip/journal/apl/79/10/10.1063/1.1401089}%
  \BibitemOpen
  \bibfield  {author} {\bibinfo {author} {\bibfnamefont {M.~K.}\ \bibnamefont
  {Sunkara}}, \bibinfo {author} {\bibfnamefont {S.}~\bibnamefont {Sharma}},
  \bibinfo {author} {\bibfnamefont {R.}~\bibnamefont {Miranda}}, \bibinfo
  {author} {\bibfnamefont {G.}~\bibnamefont {Lian}}, \ and\ \bibinfo {author}
  {\bibfnamefont {E.~C.}\ \bibnamefont {Dickey}},\ }\href {\doibase
  http://dx.doi.org/10.1063/1.1401089} {\bibfield  {journal} {\bibinfo
  {journal} {Applied Physics Letters}\ }\textbf {\bibinfo {volume} {79}},\
  \bibinfo {pages} {1546} (\bibinfo {year} {2001})}\BibitemShut {NoStop}%
\bibitem [{\citenamefont {Stillinger}(1999)}]{PhysRevE.59.48}%
  \BibitemOpen
  \bibfield  {author} {\bibinfo {author} {\bibfnamefont {F.~H.}\ \bibnamefont
  {Stillinger}},\ }\href {\doibase 10.1103/PhysRevE.59.48} {\bibfield
  {journal} {\bibinfo  {journal} {Phys. Rev. E}\ }\textbf {\bibinfo {volume}
  {59}},\ \bibinfo {pages} {48} (\bibinfo {year} {1999})}\BibitemShut {NoStop}%
\bibitem [{\citenamefont
  {Hellenbrandt}(2004)}]{doi:10.1080/08893110410001664882}%
  \BibitemOpen
  \bibfield  {author} {\bibinfo {author} {\bibfnamefont {M.}~\bibnamefont
  {Hellenbrandt}},\ }\href {\doibase 10.1080/08893110410001664882} {\bibfield
  {journal} {\bibinfo  {journal} {Crystallography Reviews}\ }\textbf {\bibinfo
  {volume} {10}},\ \bibinfo {pages} {17} (\bibinfo {year} {2004})},\ \Eprint
  {http://arxiv.org/abs/http://dx.doi.org/10.1080/08893110410001664882}
  {http://dx.doi.org/10.1080/08893110410001664882} \BibitemShut {NoStop}%
\bibitem [{\citenamefont {Curtarolo}\ \emph {et~al.}(2012)\citenamefont
  {Curtarolo}, \citenamefont {Setyawan}, \citenamefont {Hart}, \citenamefont
  {Jahnatek}, \citenamefont {Chepulskii}, \citenamefont {Taylor}, \citenamefont
  {Wang}, \citenamefont {Xue}, \citenamefont {Yang}, \citenamefont {Levy},
  \citenamefont {Mehl}, \citenamefont {Stokes}, \citenamefont {Demchenko},\
  and\ \citenamefont {Morgan}}]{Curtarolo2012218}%
  \BibitemOpen
  \bibfield  {author} {\bibinfo {author} {\bibfnamefont {S.}~\bibnamefont
  {Curtarolo}}, \bibinfo {author} {\bibfnamefont {W.}~\bibnamefont {Setyawan}},
  \bibinfo {author} {\bibfnamefont {G.~L.}\ \bibnamefont {Hart}}, \bibinfo
  {author} {\bibfnamefont {M.}~\bibnamefont {Jahnatek}}, \bibinfo {author}
  {\bibfnamefont {R.~V.}\ \bibnamefont {Chepulskii}}, \bibinfo {author}
  {\bibfnamefont {R.~H.}\ \bibnamefont {Taylor}}, \bibinfo {author}
  {\bibfnamefont {S.}~\bibnamefont {Wang}}, \bibinfo {author} {\bibfnamefont
  {J.}~\bibnamefont {Xue}}, \bibinfo {author} {\bibfnamefont {K.}~\bibnamefont
  {Yang}}, \bibinfo {author} {\bibfnamefont {O.}~\bibnamefont {Levy}}, \bibinfo
  {author} {\bibfnamefont {M.~J.}\ \bibnamefont {Mehl}}, \bibinfo {author}
  {\bibfnamefont {H.~T.}\ \bibnamefont {Stokes}}, \bibinfo {author}
  {\bibfnamefont {D.~O.}\ \bibnamefont {Demchenko}}, \ and\ \bibinfo {author}
  {\bibfnamefont {D.}~\bibnamefont {Morgan}},\ }\href {\doibase
  https://doi.org/10.1016/j.commatsci.2012.02.005} {\bibfield  {journal}
  {\bibinfo  {journal} {Computational Materials Science}\ }\textbf {\bibinfo
  {volume} {58}},\ \bibinfo {pages} {218 } (\bibinfo {year}
  {2012})}\BibitemShut {NoStop}%
\bibitem [{\citenamefont {Giannozzi}\ \emph {et~al.}(2009)\citenamefont
  {Giannozzi}, \citenamefont {Baroni}, \citenamefont {Bonini}, \citenamefont
  {Calandra}, \citenamefont {Car}, \citenamefont {Cavazzoni}, \citenamefont
  {Ceresoli}, \citenamefont {Chiarotti}, \citenamefont {Cococcioni},
  \citenamefont {Dabo}, \citenamefont {{Dal Corso}}, \citenamefont
  {de~Gironcoli}, \citenamefont {Fabris}, \citenamefont {Fratesi},
  \citenamefont {Gebauer}, \citenamefont {Gerstmann}, \citenamefont
  {Gougoussis}, \citenamefont {Kokalj}, \citenamefont {Lazzeri}, \citenamefont
  {Martin-Samos}, \citenamefont {Marzari}, \citenamefont {Mauri}, \citenamefont
  {Mazzarello}, \citenamefont {Paolini}, \citenamefont {Pasquarello},
  \citenamefont {Paulatto}, \citenamefont {Sbraccia}, \citenamefont {Scandolo},
  \citenamefont {Sclauzero}, \citenamefont {Seitsonen}, \citenamefont
  {Smogunov}, \citenamefont {Umari},\ and\ \citenamefont
  {Wentzcovitch}}]{QE-2009}%
  \BibitemOpen
  \bibfield  {author} {\bibinfo {author} {\bibfnamefont {P.}~\bibnamefont
  {Giannozzi}}, \bibinfo {author} {\bibfnamefont {S.}~\bibnamefont {Baroni}},
  \bibinfo {author} {\bibfnamefont {N.}~\bibnamefont {Bonini}}, \bibinfo
  {author} {\bibfnamefont {M.}~\bibnamefont {Calandra}}, \bibinfo {author}
  {\bibfnamefont {R.}~\bibnamefont {Car}}, \bibinfo {author} {\bibfnamefont
  {C.}~\bibnamefont {Cavazzoni}}, \bibinfo {author} {\bibfnamefont
  {D.}~\bibnamefont {Ceresoli}}, \bibinfo {author} {\bibfnamefont {G.~L.}\
  \bibnamefont {Chiarotti}}, \bibinfo {author} {\bibfnamefont {M.}~\bibnamefont
  {Cococcioni}}, \bibinfo {author} {\bibfnamefont {I.}~\bibnamefont {Dabo}},
  \bibinfo {author} {\bibfnamefont {A.}~\bibnamefont {{Dal Corso}}}, \bibinfo
  {author} {\bibfnamefont {S.}~\bibnamefont {de~Gironcoli}}, \bibinfo {author}
  {\bibfnamefont {S.}~\bibnamefont {Fabris}}, \bibinfo {author} {\bibfnamefont
  {G.}~\bibnamefont {Fratesi}}, \bibinfo {author} {\bibfnamefont
  {R.}~\bibnamefont {Gebauer}}, \bibinfo {author} {\bibfnamefont
  {U.}~\bibnamefont {Gerstmann}}, \bibinfo {author} {\bibfnamefont
  {C.}~\bibnamefont {Gougoussis}}, \bibinfo {author} {\bibfnamefont
  {A.}~\bibnamefont {Kokalj}}, \bibinfo {author} {\bibfnamefont
  {M.}~\bibnamefont {Lazzeri}}, \bibinfo {author} {\bibfnamefont
  {L.}~\bibnamefont {Martin-Samos}}, \bibinfo {author} {\bibfnamefont
  {N.}~\bibnamefont {Marzari}}, \bibinfo {author} {\bibfnamefont
  {F.}~\bibnamefont {Mauri}}, \bibinfo {author} {\bibfnamefont
  {R.}~\bibnamefont {Mazzarello}}, \bibinfo {author} {\bibfnamefont
  {S.}~\bibnamefont {Paolini}}, \bibinfo {author} {\bibfnamefont
  {A.}~\bibnamefont {Pasquarello}}, \bibinfo {author} {\bibfnamefont
  {L.}~\bibnamefont {Paulatto}}, \bibinfo {author} {\bibfnamefont
  {C.}~\bibnamefont {Sbraccia}}, \bibinfo {author} {\bibfnamefont
  {S.}~\bibnamefont {Scandolo}}, \bibinfo {author} {\bibfnamefont
  {G.}~\bibnamefont {Sclauzero}}, \bibinfo {author} {\bibfnamefont {A.~P.}\
  \bibnamefont {Seitsonen}}, \bibinfo {author} {\bibfnamefont {A.}~\bibnamefont
  {Smogunov}}, \bibinfo {author} {\bibfnamefont {P.}~\bibnamefont {Umari}}, \
  and\ \bibinfo {author} {\bibfnamefont {R.~M.}\ \bibnamefont {Wentzcovitch}},\
  }\href {http://www.quantum-espresso.org} {\bibfield  {journal} {\bibinfo
  {journal} {Journal of Physics: Condensed Matter}\ }\textbf {\bibinfo {volume}
  {21}},\ \bibinfo {pages} {395502 (19pp)} (\bibinfo {year}
  {2009})}\BibitemShut {NoStop}%
\bibitem [{\citenamefont {Mujica}\ \emph {et~al.}(2015)\citenamefont {Mujica},
  \citenamefont {Pickard},\ and\ \citenamefont {Needs}}]{mujica2015low}%
  \BibitemOpen
  \bibfield  {author} {\bibinfo {author} {\bibfnamefont {A.}~\bibnamefont
  {Mujica}}, \bibinfo {author} {\bibfnamefont {C.~J.}\ \bibnamefont {Pickard}},
  \ and\ \bibinfo {author} {\bibfnamefont {R.~J.}\ \bibnamefont {Needs}},\
  }\href@noop {} {\bibfield  {journal} {\bibinfo  {journal} {Physical Review
  B}\ }\textbf {\bibinfo {volume} {91}},\ \bibinfo {pages} {214104} (\bibinfo
  {year} {2015})}\BibitemShut {NoStop}%
\bibitem [{\citenamefont {Zhao}\ \emph {et~al.}(2012)\citenamefont {Zhao},
  \citenamefont {Tian}, \citenamefont {Dong}, \citenamefont {Li}, \citenamefont
  {Wang}, \citenamefont {Wang}, \citenamefont {Zhong}, \citenamefont {Xu},
  \citenamefont {Yu}, \citenamefont {He}, \citenamefont {Wang}, \citenamefont
  {Ma},\ and\ \citenamefont {Tian}}]{doi:10.1021/ja304380p}%
  \BibitemOpen
  \bibfield  {author} {\bibinfo {author} {\bibfnamefont {Z.}~\bibnamefont
  {Zhao}}, \bibinfo {author} {\bibfnamefont {F.}~\bibnamefont {Tian}}, \bibinfo
  {author} {\bibfnamefont {X.}~\bibnamefont {Dong}}, \bibinfo {author}
  {\bibfnamefont {Q.}~\bibnamefont {Li}}, \bibinfo {author} {\bibfnamefont
  {Q.}~\bibnamefont {Wang}}, \bibinfo {author} {\bibfnamefont {H.}~\bibnamefont
  {Wang}}, \bibinfo {author} {\bibfnamefont {X.}~\bibnamefont {Zhong}},
  \bibinfo {author} {\bibfnamefont {B.}~\bibnamefont {Xu}}, \bibinfo {author}
  {\bibfnamefont {D.}~\bibnamefont {Yu}}, \bibinfo {author} {\bibfnamefont
  {J.}~\bibnamefont {He}}, \bibinfo {author} {\bibfnamefont {H.-T.}\
  \bibnamefont {Wang}}, \bibinfo {author} {\bibfnamefont {Y.}~\bibnamefont
  {Ma}}, \ and\ \bibinfo {author} {\bibfnamefont {Y.}~\bibnamefont {Tian}},\
  }\href {\doibase 10.1021/ja304380p} {\bibfield  {journal} {\bibinfo
  {journal} {Journal of the American Chemical Society}\ }\textbf {\bibinfo
  {volume} {134}},\ \bibinfo {pages} {12362} (\bibinfo {year} {2012})},\
  \bibinfo {note} {pMID: 22803841},\ \Eprint
  {http://arxiv.org/abs/http://dx.doi.org/10.1021/ja304380p}
  {http://dx.doi.org/10.1021/ja304380p} \BibitemShut {NoStop}%
\bibitem [{Note1()}]{Note1}%
  \BibitemOpen
  \bibinfo {note} {This observation might suggest that our calculations
  overestimate the volume of P6/mmm, which is typically cited to be between 13
  to 14 {\r A}$^{3}$ per atom. However, that is the equilibrium volume under
  pressure, which will generally be smaller than the volume at ambient
  conditions}\BibitemShut {NoStop}%
\bibitem [{\citenamefont {Pickard}\ and\ \citenamefont
  {Needs}(2011)}]{0953-8984-23-5-053201}%
  \BibitemOpen
  \bibfield  {author} {\bibinfo {author} {\bibfnamefont {C.~J.}\ \bibnamefont
  {Pickard}}\ and\ \bibinfo {author} {\bibfnamefont {R.~J.}\ \bibnamefont
  {Needs}},\ }\href {http://stacks.iop.org/0953-8984/23/i=5/a=053201}
  {\bibfield  {journal} {\bibinfo  {journal} {Journal of Physics: Condensed
  Matter}\ }\textbf {\bibinfo {volume} {23}},\ \bibinfo {pages} {053201}
  (\bibinfo {year} {2011})}\BibitemShut {NoStop}%
\bibitem [{USG(2017)}]{USGS}%
  \BibitemOpen
  \href@noop {} {\enquote {\bibinfo {title} {{U.S. Geological Survey Mineral
  Commodity Summaries, 2017}},}\ }\bibinfo {howpublished}
  {\url{https://minerals.usgs.gov/minerals/pubs/commodity/diamond/mcs-2017-diamo.pdf}}
  (\bibinfo {year} {2017}),\ \bibinfo {note} {accessed: 2017-07-18}\BibitemShut
  {NoStop}%
\end{thebibliography}


%
%
\end{document}